\documentstyle[12pt]{article}
\def\np    { Nucl. Phys. }

\def\pl    { Phys. Lett. }

\def\begineq{\begin{equation}}
\def\endeq{\end{equation}}

\def\beqa{\begin{eqnarray}}
\def\eeqa{\end{eqnarray}}

\def\parn              {  \par\noindent }
\def\parbigskip        {  \par\bigskip  }

\def\parbigskipn        {  \par\bigskip\noindent  }

\def\parsmallskipn      {  \par\smallskip\noindent  }
\def\papertitlepage{\baselineskip 3.5ex \thispagestyle{empty}}
\def\Title#1{\vspace{2.5cm}\begin{center}
 {\LARGE\bf #1} \end{center} 
\vspace{2cm}}
\def\Authors#1{\begin{center} {\Large\it #1} \end{center}}
\def\Abstract{\vspace{2.7cm}\begin{center} {\Large\bf Abstract} 
           \end{center} \parbigskip}
\def\ICRRnumber#1#2{\hfill \begin{minipage}{4cm} #1
              \parn #2 \end{minipage}}

\begin{document}
\papertitlepage
\vspace*{-1 cm}
\ICRRnumber{ICRR-420-98-16}{May 1998}
\Title{The Spacetime Superalgebras   \\ 
\vskip 1.5ex in a Massive IIA Background via Brane Probes} 
\Authors{{\sc\  Takeshi Sato
\footnote{tsato@icrr.u-tokyo.ac.jp}} \\
 \vskip 3ex
 Institute for Cosmic Ray Research, University of Tokyo, \\
 Tanashi, Tokyo 188-8502 Japan \\
  }
\Abstract
We derive the spacetime superalgebras explicitly
from ``test'' p-brane actions 
in a D-8-brane (i.e. a {\it massive IIA})
background to the lowest order in $\theta$ 
via canonical formalism,
and show that the forms of the superalgebras are the same
as those in all the other D-brane (i.e. massless IIA)
backgrounds,
that is, they are indifferent to the presence
of the Chern-Simons terms
which are proportional to the mass and added to the D-brane actions
in the case of massive IIA backgrounds.
Thus, we can say that
all the D-brane background solutions including a D-8-brane are
on equal footing from the viewpoint of the superalgebras
via brane probes.
We also deduce from the algebra all the
previously known 1/4 supersymmetric
intersections of a p-brane or a fundamental string with a D-8-brane,
as the supersymmetric ``gauge fixings'' of the test branes
in the D-8-brane background.

\newpage

%

\baselineskip=0.6cm


\section{Introduction}

%

The M-theory is currently a most hopeful candidate
for a unified
theory of particle interactions\cite{tow2}\cite{wit1} and is
extensively studied\cite{tow3}\cite{sus1}.
In the literature it has been conjectured
that the M-theory is the strong coupling limit
of the type IIA string theory
and the type IIA string is obtained from it
by a dimensional reduction.
So, based on the conjecture,
the 10-dimensional type IIA theory 
and all of its constituents should
have their 11-dimensional interpretations.
As is well-known, however,
the 11-dimensional origins of
the 10-dimensional massive IIA supergravity\cite{rom1}
and its D-8-brane solutions\cite{pol3}\cite{berg3} have not been
understood completely, although some proposal have been given
to the former\cite{how4}\cite{berg4}\cite{loz1}\footnote{
Although a mass term and a cosmological one are given,
there is no static supersymmetric
D-8-brane solution in the theory proposed in
ref.\cite{how4} because of the ``wrong'' sign of the mass term,
and
an abelian isometry is assumed
in 11-dimensions in ref.\cite{berg4}.}
and the existence of M-9-brane
is conjectured for the
latter\cite{berg3}\cite{berg8}\cite{hul1}(and recently \cite{berg9}).
Briefly speaking, this problem is due to the fact that
the massive IIA supergravity has nonzero cosmological constant
proportional to the square of the mass,
which is against the no-go theorem
presented in ref.\cite{des1} 
that there is no cosmological constant in 11-dimensional supergravity.
The mass parameter can be interpreted
as the dual of a R-R 9-form\cite{pol1}\cite{berg3}
which couples only to D-8-branes.
In other words, the existence of D-8-branes
leads to a ``massive'' background.
Thus, the background including some D-8-branes 
is worth examining in order to understand the 11-dimensional origins
of not only the D-8-brane itself
but also the 10-dimensional massive IIA supergravity.

In this paper we investigate a D-8-brane background from the viewpoint 
of superalgebra.
In fact the explicit computations of superalgebras from
M-branes and D-branes had been done only for branes in the {\it flat}
background\cite{asc1}\cite{tow5}\cite{ham1}\cite{hatsu1} until
recently.
The flat background, however, leads to the {\it massless} IIA theory
because the ``flat'' spacetime implies
the vanishing of cosmological constant and hence
$m=0$. In the previous paper\cite{sato1}, however, we have proposed
the method of deriving
superalgebras explicitly from {\it branes in supersymmetric
brane backgrounds} in terms of M-theory.\footnote{
The possibility of this computation has been pointed out
in ealiar paper ref.\cite{asc1} for a different
purpose(related to nontrivial topologies),
although it is not shown explicitly there.}
We call them superalgebras via brane probes
because the test branes ``probe'' the supersymmetries which the
backgrounds possess.
The idea presented in ref.\cite{sato1} is as follows:
suppose we float a ``test'' brane as a probe in a certain brane
background which have some portions of supersymmetry.
The test brane action is initially invariant under
local super-transformation. Then, if we take the background to be the
brane solution, it is natural for the system and hence the action
to have the corresponding
portions of supersymmetry bacause of the two reasons:
first, the test brane is assumed to be so light that it does not
affect the background. Second, since the configuration
of the test brane
including its orientation has not yet been fixed at this moment,
it does not break any supersymmetry.
Thus, we can define the corresponding Noether supercharge expressed
in terms of the
worldvolume fields and their conjugate momenta,
and compute its commutator to obtain the superalgebra.
The reliability of this method
has been confirmed by deducing from the algebras
all the 1/4-supersymmetric
intersections of various combinations of
two M-branes (and some of bound states of them)
known
before\cite{guv1}\cite{tow9}\cite{gau1}\cite{tow6}\cite{izq1},
as the supersymmetric ``gauge fixings'' of the test
branes in the background branes.
Now, we apply the method to the 10-dimensional massive IIA case.

There are two aims in this paper:
the first is to investigate the contributions 
to the superalgebras
from
the Chern-Simons terms of the worldvolume
vector field which are proportional to
the mass and added to the brane actions
in the cases of
massive IIA backgrounds\cite{berg6}\cite{berg7}\cite{berg5}.
If there were their contributions,
it might be significant to know them
in the following two points:
the one is that
the contributions might constrain the conjectured M-9-brane background 
so that the dimensional reduction of
the superalgebras in an M-9-brane background should give rise to the
contributions.
The other is that the terms might exhibit
the possibility of the existence of some new
BPS worldvolume solitons.
We note that there is no other way but to take the background to be a
certain massive IIA (i.e. not flat) one
in order to get the contributions of the CS terms
to the algebras.
The second aim is to obtain explicitly the supersymmetric
intersections known before from the algebras as the supersymmetric
``gauge fixings'' of
the test branes in the background D-8-brane.

The concrete procedures are as follows\cite{sato1}: 
we will take the background to be a (1/2-supersymmetric)
D-8-brane solution
which actually consists of
such a large number of coincident D-8-branes
that our ``test brane'' approximation can be justified, and
construct the superfields in terms of the component fields of the
massive IIA supergravity in the D-8-brane background.\footnote{In 
fact we need to construct the expressions of the
superfields with mass corrections to the usual ones.
Thus, we construct them in this specific background. 
Those in {\it general} massive IIA backgrounds will be presented in
ref.\cite{sato3}.}
Then, we substitute the solution for the test brane action as was done
in ref.\cite{tow8}, check the invariance of
the action under the unbroken supersymmetry
transformation, derive the representation of the supercharge 
in terms of the worldvolume fields of the
test brane and their conjugate momenta, and
compute its commutator as the superalgebra.

We note that our computations are performed only up to the low
orders in $\theta$ which might contribute to the the central charges 
at zeroth order in $\theta$ because they suffice
for the aims stated above.
And in the same way as in the previous paper\cite{sato1},
we reduce the superspace with the supercoordinates $(x,\theta)$ 
to that with coordinates $(x,\theta^{+})$ (or $(x,\theta^{-})$)
where the sign $+$($-$) of $\theta^{+}$($\theta^{-}$)
implies that it has a positive (negative)
``chirality'' determined by the background. 
The reason is the following: 
since half of supersymmetry is already not the symmetry of the system 
owing to the existence of the background brane,
the corresponding
parameter $\theta^{-} $($\theta^{+} $)
must not be transformed.
Thus, the conjugate momentum of $\theta^{-} $($\theta^{+} $)
does not appear in the
supercharge $Q^{+}$($Q^{-}$), which means that
the terms including $\theta^{-}$($\theta^{+} $)
does not contribute to the
central charges at zeroth order in $\theta$.
Therefore, we set $\theta^{-}=0$($\theta^{+}=0 $)
from the beginning.

The consequence is that the superalgebras in the
D-8-brane background have the same forms as those
not only in the flat
but also in all the other D-brane (i.e. massless IIA) backgrounds
except that half of the supercharges are projected out
in the cases of brane backgrounds.
In other words, they are regardless of the Chern-Simons terms
which are added to the brane actions
in the case of massive backgrounds.
Thus, we can say that
{\it all the D-brane background solutions including a D-8-brane are
on equal footing from the viewpoint of the superalgebras
via brane probes}.
In addition we can also deduce explicitly from the superalgebras 
all the 1/4-supersymmetric intersections 
of all the D-branes and a fundamental string
with a D-8-brane known before, as in the M-brane
cases\cite{sato1}.

On the other hand,
bound states of a D-8-brane with other branes
cannot be discussed
on the basis of the algebras, and it is unclear
whether or not the test branes have boundaries
near the background brane, both as opposed
to the M-brane cases\cite{sato1}\cite{tow8}.
Both of the above are due to the peculiar
behavior of the harmonic function determined by the D-8-brane
background.
We will discuss these points in section 4.

This paper is organized as follows:
in section 2 we give some preliminaries in terms of
a D-8-brane background
and the expressions of the superfields in the background.
In section 3 we derive the superalgebras from
all the D-p-brane and a fundamental string 
in the D-8-brane background.
In section 4 we give short summary and discussions.  

Before presenting our results we give the notations used
in this paper:
we use ``mostly plus'' metrics for both worldvolume and spacetime.
And we use Majorana ($32 \times 32$) representation for Gamma matrices 
$\Gamma_{\hat{m}}$ which are all real and satisfy 
$ \{ \Gamma_{\hat{m}} , \Gamma_{\hat{n}} \} 
= 2\eta_{\hat{m}\hat{n}}$. 
$\Gamma_{\hat{0}}$ is antisymmetric and others 
symmetric. Charge Conjugation is ${\cal C}=\Gamma_{\hat{0}}$.
We use capital latin letters($M,N,..$) for superspace indices, 
small latin letters($m,n,..$) for vectors and early small greek letters
($\alpha,\beta$,..) for spinors. Furthermore, we use late greek letters     
($\mu,\nu,..$) for vectors paralell to the background brane.
We use {\it hatted letters} ($\hat{M},\hat{m},\hat{\alpha}..$) 
for {\it all the inertial frame indices}
and middle latin letters($i,j,..$) for worldvolume vectors.
Finally, we set $\Gamma_{\hat{0}\hat{1}..\hat{9}}\Gamma_{11}=1$.


\section{Superspace preliminaries in a massive background}

In this section we give some preliminaries about the D-8-brane
background to be chosen and the expressions of the superfields
and the super-transformation in the background.

The D-8-brane background solution 
we choose is\cite{pol1}\cite{berg3}\footnote{
In fact the D-8-brane solution has the freedom
to choose an arbitrary function\cite{berg3}.
In this paper we choose it so that the metric saticfies the ``harmonic 
function rule'' presented in ref.\cite{tsey2}.}
\beqa
ds^{2} &=& H^{-1/2}dx^{\mu}dx^{\nu}\eta_{\mu\nu}+H^{1/2}dy^{2}\nonumber\\
e^{\phi} &=& H^{-5/4}\nonumber\\
F_{m_{1}..m_{10}}&=&-H^{-2}m\epsilon_{m_{1}..m_{10}}\label{d8sol}
\eeqa
where $\eta_{\mu\nu}$ is the 9-dimensional Minkovski metric with
coordinates $x^{\mu}$ and H is a harmonic function 
on the transverse 1-dimensional space with a coordinate $y$.
$m$ is a mass parameter which is constant and positive,
and $\epsilon_{01..9}=-1 $.
In this paper
we choose $H=m|y|$ which means that the D-8-brane lies at $y=0 $.

This background admits a Killing spinor $\varepsilon$
which satisfies 
\beqa
\delta\psi_{m}
&=&(\partial_{m}+\frac{1}{4}\omega_{m}^{\ \hat{r}\hat{s}}
\Gamma_{\hat{r}\hat{s}}
+\frac{1}{8}m e^{\phi}\Gamma_{m})
\varepsilon =0\label{suptra1}\\
\delta \lambda&=&-\frac{1}{2\sqrt{2}}(\Gamma^{m}\partial_{m}
\phi+\frac{5}{4}m e^{\phi})\varepsilon =0\label{suptra2} 
\eeqa
where the last terms in (\ref{suptra1})
and (\ref{suptra2}) are mass corrections.
Then, the Killing spinor has the form
$\varepsilon = H^{-1/8}\varepsilon_{0} $ where $\varepsilon_{0}$
has a definite chirality,
i.e. $ \Gamma_{\hat{y}}
\varepsilon_{0}=+\varepsilon_{0}$ for $y>0$ and $\Gamma_{\hat{y}}
\varepsilon_{0}=-\varepsilon_{0}$ for $y<0$.
From now on, we concentrate the discussions on those at $y>0 $
for simplicity,
although we continue them so as to convert the results
to those at $y<0 $ at any time.

Since $\Gamma_{\hat{y}}$ satisfies 
$\Gamma_{\hat{y}}^{T}=\Gamma_{\hat{y}}$ and $\Gamma_{\hat{y}}^{2}=1$,
both $\frac{1\pm\Gamma_{\hat{y}}}{2}$ and
$\frac{1\pm\Gamma_{\hat{y}}^{T}}{2}$ are projection operaters.
So, if we denote 
$\ \frac{1\pm\Gamma_{\hat{y}}}{2}\zeta$ as $\zeta^{\pm}$
for a spinor $\zeta $,
the system (i.e. the action)
is invariant under the transformation 
generated by the supercharge $Q^{+}$. We interpret that
it is due to the following two reasons:
first, the test brane is assumed to be so light that 
it does not affect the background geometry. Second,
its configuration are not fixed yet at this moment.
On the other hand, the background and the brane action 
are not invariant under the transformation by $Q^{-}$.
So, we should set the corresponding transformation parameter
$\varepsilon^{-}$ to be zero,
which means that the conjugate momentum $\Pi^{-}$ 
of $\theta^{-}$ does not appear in the Noether charge $Q^{+}$ 
only whose algebra we are interested in.
Therefore, the terms including $\theta^{-}$ {\it never} contribute to 
the central charges at zeroth order in $\theta$.
Thus, we set {\it from the beginning}
\beqa
\theta^{-}=0. \label{thetam}
\eeqa
From now on we will use these freely in all the cases we treat 
in this paper. Related with this,
we exhibit the property of 
$\bar{\Gamma}:\ \{ \bar{\Gamma},{\cal C} \}=0$. 
We note that the argument above is also true for $y<0$,
in which the sign $+$ is exchanged with $-$.

Now, we are prepared to construct the superfields and
the super-transformation in the background.
We begin with
the supervielbein $E_{M}^{\ \hat{A}}$.
Because of the last term in (\ref{suptra1}),
$E_{m}^{\ \hat{\alpha}}$ receives the mass correction 
which do not appear in the one obtained
from a dimensional reduction of $E_{M}^{\ \hat{A}}$
in 11 dimensions\cite{cre1}.
It is
\beqa
E_{m}^{\hat{\alpha}}=\psi_{m}^{\hat{\alpha}}+
[ \frac{1}{4}\omega_{m}^{\ \hat{r}\hat{s}}
(\Gamma_{\hat{r}\hat{s}}\theta)^{\hat{\alpha}}
+\frac{1}{8}m e^{\phi} 
(\Gamma_{m}\theta)^{\hat{\alpha}} ] +{\cal O}(\theta^{2}).
\eeqa
Then, the superspace 1-form basis on the inertial frame 
$E^{\hat{A}}=dZ^{M}E_{M}^{\ \hat{A}}$ is\footnote{
In fact we need to know the (vanishing of the)
contribution from $E_{m}^{\hat{n}}$  
at order $\theta^{2}$ as stated in the previous paper. 
We can infer its vanishing 
in this specific simple background.}
\beqa
E^{\hat{\mu}} & = & dx^{\nu}H^{-1/4}\delta_{\nu}^{\ \hat{\mu}}
+i\bar{\theta^{+}}\Gamma^{\hat{\mu}}d\theta^{+}
+{\cal O}(\theta^{4})\\
E^{\hat{y}} & = & dy H^{1/4}+{\cal O}(\theta^{4})\\
E^{\hat{\alpha}} & = & d\theta^{\hat{\alpha}+}+\frac{1}{8}m dy 
H^{-1}(\Gamma_{\hat{y}}\theta^{+})^{\hat{\alpha}}+{\cal O}(\theta^{3}).
\eeqa
Since 
these 1-forms have no (curved) superspace indices,
they are invariant
under the super-coordinate transformation\cite{cre1}
$\delta Z^{M}=\Xi^{M}$ in this background given by
 \beqa
\Xi^{\mu} &=&i\bar{\varepsilon}^{+}\Gamma^{\mu}\theta^{+} 
+ {\cal O}(\theta^{3}) \nonumber \\
\Xi^{y} &=& 0+ {\cal O}(\theta^{3})  \nonumber \\
\Xi^{\alpha} &=& \varepsilon^{\alpha +}+ {\cal O}(\theta^{2}).
\label{supertr}
\eeqa
We can easily check the invariance explicitly
up to second order in $\theta $.
Note that the coordinate $y$ transverse to the D-8-brane is not
transformed, which means that this transformation
corresponds to the supertranslation symmetry paralell to the
background D-8-brane. Thus, we can define the corresponding Noether
supercharge.

Next, we will construct the superspace gauge potentials.
The NS-NS 2-form $B_{2}$ in the background (\ref{d8sol})
is introduced
by the 3-form field strength\cite{berg5}
\beqa
H_{3}\equiv dB_{2}=E^{\hat{m}}E^{\hat{\alpha}}E^{\hat{\beta}}
(\Gamma_{11}\Gamma_{m})_{\hat{\alpha}\hat{\beta}}.
\eeqa

Here, we make an assumption
that all the supersymmetric cocycles 
in the superspace (but not those in the bosonic space)
in the background 
are trivial.
Then, since $H_{3}$ is invariant under (\ref{supertr}),
it holds
$\delta B_{2}=d(\bar{\varepsilon}^{+}\Delta'^{(1)})$ for a 1-form
$\Delta'^{(1)}$, or explicitly,
\footnote{Although
$\hat{\alpha}$ of $\theta^{\hat{\alpha}+}$
is the index of the inertial frame, 
$\theta^{\hat{\alpha}}
=\theta^{\beta}\delta_{\beta}^{\hat{\alpha}}+{\cal O}(\theta^{3}) $.
So, we need not distinguish the two indices in this paper.} 
\beqa
\delta B_{2}\equiv d (\bar{\varepsilon}^{+}\Delta'^{(1)})=
d(-idyH^{1/4}\bar{\varepsilon}^{+}\Gamma_{11}
\Gamma_{\hat{y}}\theta^{+}
+{\cal O}(\theta^{3})).\label{sptrans2}
\eeqa

The remaining fields are the R-R r-form superspace gauge potentials
$C^{(r)}$ (r=1,3,5,7,9). We introduce a R-R potential $C$
as a formal sum of odd form potentials 
$C= \sum_{r=1}^{'9} C^{(r)}$, whose field strength is\cite{berg7}
\beqa
R(C) \equiv dC-H_{3}C+ m e^{B}\label{rrfs1}
\eeqa
where $R(C)$ is expressed as a formal sum of even form field strengths
$R(C)=\sum_{n=2}^{''10} R^{(n)}$.

The superspace constraints on the components of these field strengths 
with spinor indices in bosonic backgrounds are\cite{berg3}
\beqa
R_{\hat{\alpha}\hat{\beta}
\hat{m_{1}}..\hat{m_{n}}} = 
\left\{
  \begin{array}{@{\,}ll} 
     i e^{-\phi}(\Gamma_{\hat{m_{1}}..\hat{m_{n}}}\Gamma_{11}
)_{\hat{\alpha}\hat{\beta}} & (n=0,4,8)\\
 i e^{-\phi}(\Gamma_{\hat{m_{1}}..\hat{m_{n}}}
)_{\hat{\alpha}\hat{\beta}} & (n=2,6)
  \end{array}
\right.\label{rrfs2}
\eeqa
and the others vanish.
The only non-vanishing bosonic component in the background
(\ref{d8sol}) is 
\beqa
R^{(10)}_{\hat{m_{1}}..\hat{m_{10}}}=
-H^{-2}m\epsilon_{\hat{m_{1}}..\hat{m_{10}}}\label{rrfs3}
\eeqa
where $\epsilon_{01..9}=-1$. From (\ref{rrfs1}),
(\ref{rrfs2}) and (\ref{rrfs3}) we get the expressions of $C^{(r)}$.
We do not present them except for $C^{(9)}$ because the others are
rather trivial. $C^{(9)}$ is given by 
\beqa
C^{(9)}&=&\frac{1}{9!}H^{-2}mydx^{\nu_{1}}..dx^{\nu_{9}}
(-\epsilon_{\nu_{1}..\nu_{9}y})\nonumber\\
& &+\frac{i}{8!}H^{-3/4}dx^{\nu_{1}}...dx^{\nu_{8}}
\delta_{\nu_{1}}^{\mu_{1}}..\delta_{\nu_{8}}^{\mu_{8}}
\bar{\theta}^{+}\Gamma_{\hat{\mu_{1}}..\hat{\mu_{8}}}
\Gamma_{11}d\theta^{+}
+{\cal O}(\theta^{4}).\label{c9}
\eeqa

Next, we will discuss the properties of
the super-transformations of $C^{(r)}$.
In the first place, from the invarince of $R^{(2)}$ in (\ref{rrfs1}),
we get
\beqa
0=d\delta C^{(1)}+m\delta B_{2}
=d(\delta C^{(1)}+m\bar{\varepsilon}^{+}\Delta'^{(1)})\label{delb2}.
\eeqa
And from the invarince of $R^{(4)}$ in (\ref{rrfs1}) and by using
(\ref{delb2}), we can show that
$d(\delta C^{(3)}-\delta C^{(1)}B_{2})=0$.
Repeating similar steps,
we can show that $\delta C^{(r)}$ plus some terms
are closed. By using the formal sum of odd forms C, it is written as
\beqa
d(\delta C e^{-B_{2}}+m\bar{\varepsilon}^{+}\Delta'^{(1)})=0.
\eeqa
From this property and the triviality of the supersymmetric
cocycles in the superspace,
each of the odd r-forms of $\delta C e^{-B_{2}}
+m\bar{\varepsilon}^{+}\Delta'^{(1)} $ can be written 
as a d-exact form given by
\beqa
\delta C^{(1)}+\bar{\varepsilon}^{+}\Delta'^{(1)}
&\equiv& d(\bar{\varepsilon}^{+}\Delta^{(0)})
=d(-iH^{5/4}\bar{\varepsilon}^{+}
\Gamma_{11}\theta^{+}+{\cal O}(\theta^{3}))\label{sptrarr1}
\\
\delta C^{(3)}-\delta C^{(1)}B_{2}
&\equiv& d(\bar{\varepsilon}^{+}\Delta^{(2)})
=d(-iH^{5/4}dx^{\nu}\delta_{\nu}^{\hat{\mu}}dy
\bar{\varepsilon}^{+}\Gamma_{\hat{\mu}}\Gamma_{\hat{y}}\theta^{+}
+{\cal O}(\theta^{3})\label{sptrarr3})
\\
\delta C^{(5)}-\delta C^{(3)}B_{2}&+&\frac{1}{2}\delta C^{(1)}(B_{2})^2
\equiv d(\bar{\varepsilon}^{+}\Delta^{(4)})\nonumber\\
&=& d(\frac{i}{4!}H^{1/4}dx^{\nu_{1}}...dx^{\nu_{4}}
\delta_{\nu_{1}}^{\hat{\mu_{1}}}...\delta_{\nu_{4}}^{\hat{\mu_{4}}}
\bar{\varepsilon}^{+}\Gamma_{\hat{\mu_{1}}...\hat{\mu_{4}}}
\Gamma_{11}\theta^{+}
+{\cal O}(\theta^{3}))\label{sptrarr5}
\\
\delta C^{(7)}-\delta C^{(5)}B_{2}&+&\frac{1}{2}\delta C^{(3)}(B_{2})^2
-\frac{1}{3!}\delta C^{(1)}(B_{2})^3
\equiv d(\bar{\varepsilon}^{+}\Delta^{(6)})\nonumber\\
&=&d(\frac{i}{5!}H^{1/4}dx^{\nu_{1}}...dx^{\nu_{5}}
\delta_{\nu_{1}}^{\hat{\mu_{1}}}...\delta_{\nu_{5}}^{\hat{\mu_{5}}}
dy\delta_{y}^{\hat{y}}
\bar{\varepsilon}^{+}\Gamma_{\hat{\mu_{1}}...\hat{\mu_{5}}\hat{y}}
\theta^{+}
+{\cal O}(\theta^{3}))\label{sptrarr7}
\\
\delta C^{(9)}-\delta C^{(7)}B_{2}&+&
\frac{1}{2}\delta C^{(5)}(B_{2})^2
-\frac{1}{3!}\delta C^{(3)}(B_{2})^3
+\frac{1}{4!}\delta C^{(1)}(B_{2})^4
\equiv d(\bar{\varepsilon}^{+}\Delta^{(8)})\nonumber\\
&=&0+{\cal O}(\theta^{3}).\label{sptrarr9}
\eeqa
We note that $B_{2}$ in (\ref{sptrarr3})-(\ref{sptrarr9})
does not contribute to $\Delta^{(r)}|_{(\theta^{+})^1}$ 
for $r=2,4,6,8$ because
$B_{2}$ is ${\cal O}(\theta^{2})$.
Thus,
$C^{(9)}$ is invariant at least
up to second order in $\theta$.
We use the identity $\Gamma_{\hat{0}\hat{1}..\hat{9}}\Gamma_{11}=1$
to show this property. We also note that this is also true
for the area $y<0 $ because the equation
$my\Gamma_{\hat{y}}\theta=H\theta$ holds for both ranges of $y$.

\section{The superalgebras from all the
D-branes and a fundamental string in the massive IIA background}

In this section we will discuss
the superalgebras from all the
D-p-branes and a fundamental string in the 
D-8-brane background. We assume throughout this section
that the test branes are within the
region $y>0$.

The super D-p-brane action in a massive IIA background
takes the form\cite{berg6}\cite{berg7} 
\beqa
S_{p} &=& S_{p}^{(0)}+S_{p}^{WZ}\nonumber\\ 
&=& -\int d^{p+1}\xi e^{-\phi}\sqrt{-det(g_{ij}+{\cal F}_{ij})}
+\int (C e^{{\cal F}}|_{{\rm p-form}}+\frac{m}{(p/2+1)!}
V(dV)^{p/2})\label{action}
\eeqa
where $g_{ij}=E_{i}^{\hat{m}}E_{j}^{\hat{n}}\eta_{\hat{m}\hat{n}}$ is
the induced worldvolume metric.
$E_{i}^{\ \hat{m}}=\partial_{i}Z^{M}
E_{M}^{\hat{m}}$ and
${\cal F}_{ij}$ are the components of a modified worldvolume 2-form
field strength
\beqa
{\cal F}=F-B_{2}
\eeqa
where $F=dV$ is the field strength of worldvolume 1-form $V$.
$C$ is the sum of the worldvolume r-forms
induced by the superspace r-form gauge potentials.
In the second term (p+1)-forms are
chosen among the exterior products of 
the forms.
The last term in (\ref{action}) is the Chern-Simons term
which is added in the case of {\it massive} IIA backgrounds.

The super-transformation of the worldvolume 1-form $V$ is determined
by the requirement of the invariance of the modified worldvolume 2-form
field strength ${\cal F} $ in order that the action be invariant under 
the transformation (\ref{supertr}). Then, the transformation of it is
\beqa
\delta V=\bar{\varepsilon}^{+}\Delta'^{(1)}
=-idy H^{1/4}\bar{\varepsilon}^{+}
\Gamma_{11}\Gamma_{\hat{y}}\theta^{+}.\label{sptrav}
\eeqa

As the super-transformations of all the fields have been given,
we will check the invariance of the action (\ref{action}) in the next.
Since $E_{i}^{\ \hat{m}}$ and ${\cal F}_{ij}$
are invariant under the super-transformation(\ref{supertr}),
so are the Born-Infeld
actions $S^{(0)}_{p}$ 
for all the D-p-branes.
On the other hand, the Wess-Zumino terms ${\cal L}_{p}^{WZ}$
are transformed as
\beqa
\delta{\cal L}_{p}^{WZ}= 
\left\{
  \begin{array}{@{\,}ll} 
    d(\bar{\varepsilon}^{+}
      \Delta^{(0)}) & {\rm (for\  0-brane)}\\
    d[ (\bar{\varepsilon}^{+}\Delta e^{dV})|_{{\rm p-form}}
       +\frac{p/2}{(p/2+1)!}m\delta V V(dV)^{p/2-1}
        ]& {\rm (for \ 2,4,6,8-brane)}
  \end{array}
\right.\label{sptrawz}
\eeqa
where $\Delta$ is introduced as a formal sum of even forms
$\Delta \equiv \Sigma_{r=0}^{'8} \Delta^{(r)}$.
Thus, the actions are invariant up to total derivative
under (\ref{supertr}) to {\it full} order in $\theta$,
and we can define the Noether supercharges
$Q_{\alpha}^{+}$ in the Hamiltonian formulation
as integrals over the test branes at fixed time ${\cal M}_{p} $,
which take the forms\cite{asc1}
\beqa
Q_{\alpha}^{+}\equiv
\left\{
  \begin{array}{@{\,}ll} 
    Q_{\alpha}^{(0)+}-i({\cal C}\Delta^{(0)})_{\alpha}
      & {\rm (for\  0-brane)}\\
    Q_{\alpha}^{(0)+}-i\int_{{\cal M}_{p}}
             [ \{({\cal C}\Delta e^{dV})
             |_{{\rm p}}\}_{\alpha}
       -i\frac{p/2}{(p/2+1)!}m({\cal C}
       \Delta^{'(1)})_{\alpha} V(dV)^{p/2-1}]
        & {\rm (for \ 2,4,6,8brane).}
  \end{array}
\right.\label{spcharge}
\eeqa 
$Q_{\alpha}^{(0)+}$  are the contributions from $S_{p}^{(0)} $
given by
\beqa
Q_{\alpha}^{(0)+}=
\left\{
  \begin{array}{@{\,}ll} 
    i\Pi_{\alpha}^{+}-\Pi_{\mu}
      ({\cal C}\Gamma^{\mu}\theta^{+})_{\alpha}
      & {\rm (for\  0-brane)}\\
    \int_{{\cal M}_{p}}[i\Pi_{\alpha}^{+}-\Pi_{\mu}
      ({\cal C}\Gamma^{\mu}\theta^{+})_{\alpha}
    +i{\cal P}^{\underline{i}}
     (-i\partial_{\underline{i}}y H^{1/4}
\Gamma_{11}\Gamma_{\hat{y}}\theta^{+})]
        & {\rm (for \ 2,4,6,8brane)}
  \end{array}
\right.\label{spcharge0}
\eeqa
where $\underline{i}$ is the space index of test D-branes and
$\Pi_{\alpha}^{+} $, $\Pi_{\mu} $ and ${\cal P}^{\underline{i}}$
are the conjugate momenta of $\theta^{+} $, $x^{\mu}$ and
$V_{\underline{i}} $, respectively.
We exhibit the more detailed expression of
${\cal P}^{\underline{i}}\equiv {\cal P}^{(0)\underline{i}}
+{\cal P}^{WZ\underline{i}}$ which consists of
\beqa
{\cal P}^{(0)\underline{i}}&\equiv& \frac{\delta S_{p}^{(0)}}
{\delta V{\underline{i}}}\\
{\cal P}^{WZ\underline{i}}&=&
\epsilon^{0\underline{i}j_{1}..j_{p-1}}[\{(Ce^{dV})
|_{{\rm p-form}}\}_{j_{1}..j_{p-1}}
+\frac{p/2}{(p/2+1)!}m \{V(dV)^{p/2-1}\}_{j_{1}..j_{p-1}}].
\eeqa
We note that $\Pi_{y}$ does not appear in the supercharges
because $y$ is not transformed at least up to second order
in $\theta$. We also note that the components
independent of $V$ and $\theta$ in the momentum $\Pi_{\mu}$
are originated only from the Born-Infeld actions
except for the case of a test D-8-brane,
where there is the contribution from
the bosonic potential $ C^{(9)}$ of (\ref{c9})
in the Wess-Zumino term.
From (\ref{spcharge}) and (\ref{spcharge0})
the superalgebras can be derived.

Now, we will discuss each of the D-p-branes in the D-8-brane background
separately. 

(a) a test D-0-brane in the D-8-brane background

In this case the superalgebra is obtained as
\beqa
\{ Q_{\alpha}^{+}, Q_{\beta}^{+}\} =2\Pi_{\mu}
({\cal C}\Gamma^{\mu})_{\alpha\beta}
+2H^{5/4}({\cal C}\Gamma_{11})_{\alpha\beta}
+{\cal O}(\theta^{2}).\label{spalgd0}
\eeqa
We discuss the implications of the algebra.
The last term implies that 1/4 spacetime supersymmetry is preserved
in the static gauge $\partial_{0}x^{m}=\delta_{0}^{m}$ and
$\partial_{i}x^{0}=\delta_{i}^{0} $.
This is the D-0-brane/D-8-brane ``intersection'' preserving
1/4 supersymmetry, which is deduced from (1$|$MW,M9)
by a double dimensional reduction\cite{tow6} of the string
intersection.
We note that the form of the algebra is the same as that
in the flat case\cite{ham1} except
that the supercharge is projected out.
We will discuss this point in detail in the next case.

(b) a test D-2-brane in the D-8-brane background

In this case the superalgebra is
\beqa
\{ Q_{\alpha}^{+}, Q_{\beta}^{+}\}&=& 
2\int_{{\cal M}_{2}}d^{2}\xi\ \Pi_{\mu}
({\cal C}\Gamma^{\mu})_{\alpha\beta}
+2\int_{{\cal M}_{2}}d^{2}\xi{\cal P}^{\underline{i}}
\partial_{\underline{i}}y
({\cal C}\Gamma_{y}\Gamma_{11})_{\alpha\beta}
+2\int_{{\cal M}_{2}}H^{5/4}dx^{\mu}dy
({\cal C}\Gamma_{\mu y})_{\alpha\beta}\nonumber\\
& &+2\int_{{\cal M}_{2}} H^{5/4}dV({\cal C}\Gamma_{11})_{\alpha\beta}
+2\frac{m}{2}\int_{{\cal M}_{2}} V dy 
({\cal C}\Gamma_{y}\Gamma_{11})_{\alpha\beta}
+{\cal O}(\theta^{2}).\label{spalgd2}
\eeqa
We note that the fifth term in (\ref{spalgd2})
comes from the Chern-Simons term.
If the term did not vanish, it might imply 
the existence of some new BPS worldvolume solitons.
The ${\cal P}^{WZ\underline{i}}$ in the second term, however, 
cancels out the fifth term. Therefore,
adding the Chern-Simons term
does not give rise to any new terms in the superalgebra. 
In other words, the superalgebra
in the D-8-brane background has the same form as that
in the flat background except that half of $Q_{\alpha}$ is
projected out, and so are the superalgebras from the other p-branes
in the D-8-brane background, as presented later.
We note that this is also the case
with those from branes in D-p-brane backgrounds for $p\neq 8$,
although we do not present them here.
Thus, we can say that {\it all the D-brane background solutions are
on equal footing from the viewpoint of the superalgebras
via brane probes.}

The implications of the algebra are as follows:
The third term means that the string intersection lead to the
1/4-supersymmetric configuraiton in the D-8-brane background,
i.e (1$|$2,8), which is deduced from (1$|$M2,M9) by a dimensional
reduction of the space parallel to the M-9-brane
but orthogonal to the M-2-brane\cite{tow6}.
The other terms imply the {\it possibilities} that some
supersymmetric triple intersections exist, but these configuration
{\it do not always exist}, as was known before\cite{ham1}.
For example, a central charge in
the superalgebra deduced from a D-6-brane in the flat background 
seemes to give the possibility of the existence of
D-0-brane/D-6-brane bound state, but actually
it does not exist\cite{pol4}.

For the cases of the other D-p-brane, we present only the final
results of superalgebras and their interpretations.

(c) a test D-4-brane in the D-8-brane background

The superalgebra is
\beqa
\{ Q_{\alpha}^{+}, Q_{\beta}^{+}\}&=& 
2\int_{{\cal M}_{4}}d^{4}\xi\ \Pi_{\mu}
({\cal C}\Gamma^{\mu})_{\alpha\beta}
+2\int_{{\cal M}_{4}}d^{4}\xi{\cal P}^{(0)\underline{i}}
\partial_{\underline{i}}y
({\cal C}\Gamma_{y}\Gamma_{11})_{\alpha\beta}\nonumber\\
& &+\frac{2}{4!}\int_{{\cal M}_{4}}H^{5/4}dx^{\mu_{1}}..dx^{\mu_{4}}
({\cal C}\Gamma_{\mu_{1}..\mu_{4}}\Gamma_{11})_{\alpha\beta}
+2\int_{{\cal M}_{4}}H^{5/4}dx^{\mu}dydV
({\cal C}\Gamma_{\mu y})_{\alpha\beta}\nonumber\\
& &+2\int_{{\cal M}_{4}}
H^{5/4}(dV)^{2}({\cal C}\Gamma_{11})_{\alpha\beta}
+{\cal O}(\theta^{2}).\label{spalgd4}
\eeqa
The third term means that 1/4 spacetime supersymmetry is preserved
if the test D-4-brane is parallel to any 4-dimensional subspaces of
the background D-8-brane, i.e. (4$|$4,8).
This is deduced from (5$|$M5,M9) preserving 1/4 supersymmetry
by a double dimensional reduction of the 5-brane intersection\cite{tow6}.

(d) a test D-6-brane in the D-8-brane background
\beqa
\{ Q_{\alpha}^{+}, Q_{\beta}^{+}\}= 
2\int_{{\cal M}_{6}}d^{6}\xi\ \Pi_{\mu}
({\cal C}\Gamma^{\mu})_{\alpha\beta}
&+&2\int_{{\cal M}_{6}}d^{6}\xi{\cal P}^{(0)\underline{i}}
\partial_{\underline{i}}y
({\cal C}\Gamma_{y}\Gamma_{11})_{\alpha\beta}\nonumber\\
+\frac{2}{5!}\int_{{\cal M}_{6}}H^{5/4}dx^{\mu_{1}}..dx^{\mu_{5}}dy
({\cal C}\Gamma_{\mu_{1}..\mu_{5}y})_{\alpha\beta}
&+&\frac{2}{4!}\int_{{\cal M}_{6}}H^{5/4}dx^{\mu_{1}}..dx^{\mu_{4}}dV
({\cal C}\Gamma_{\mu_{1}..\mu_{4}}\Gamma_{11})_{\alpha
\beta}\nonumber\\
+2\int_{{\cal M}_{6}}H^{5/4}dx^{\mu}dy(dV)^{2}
({\cal C}\Gamma_{\mu y})_{\alpha\beta}
&+&2\int_{{\cal M}_{6}}
H^{5/4}(dV)^{3}({\cal C}\Gamma_{11})_{\alpha\beta}
+{\cal O}(\theta^{2}).\label{spalgd6}
\eeqa
The third term means that the 5-brane intersection
of the D-6-brane with the background D-8-brane
leads to the preservation of
1/4 spacetime supersymmetry, i.e. (5$|$6,8).
This is deduced from (5$|$MKK,M9) preserving 1/4 supersymmetry
by a direct dimensional reduction in terms of
M-Kalza-Klein monopole\cite{tow6}.

(e) a test D-8-brane in the D-8-brane background
\beqa
\{ Q_{\alpha}^{+}, Q_{\beta}^{+}\}= 
2\int_{{\cal M}_{8}}d^{8}\xi\ \Pi_{\mu}
({\cal C}\Gamma^{\mu})_{\alpha\beta}
&+&2\int_{{\cal M}_{8}}d^{8}\xi{\cal P}^{(0)\underline{i}}
\partial_{\underline{i}}y
({\cal C}\Gamma_{y}\Gamma_{11})_{\alpha\beta}\nonumber\\
+\frac{2}{5!}\int_{{\cal M}_{8}}
H^{5/4}dx^{\mu_{1}}..dx^{\mu_{5}}dydV
({\cal C}\Gamma_{\mu_{1}..\mu_{5}y})_{\alpha\beta}
&+&\frac{2}{4!}\int_{{\cal M}_{8}}
H^{5/4}dx^{\mu_{1}}..dx^{\mu_{4}}(dV)^{2}
({\cal C}\Gamma_{\mu_{1}..\mu_{4}}\Gamma_{11})_{\alpha
\beta}\nonumber\\
+2\int_{{\cal M}_{8}}H^{5/4}dx^{\mu}dy(dV)^{3}
({\cal C}\Gamma_{\mu y})_{\alpha\beta}
&+&2\int_{{\cal M}_{8}}
H^{5/4}(dV)^{4}({\cal C}\Gamma_{11})_{\alpha\beta}
+{\cal O}(\theta^{2}).\label{spalgd8}
\eeqa
In this case there are no central charges which are composed
purely of the products of the form $dx^{m}$.
The momentum $ \Pi_{\mu} $, however,
includes the following two terms independent of $dV$ (and $\theta$),
given by
\beqa
\Pi_{\mu}=\Pi_{\mu}^{(0)} + \frac{1}{8!}H^{-2}my
\epsilon^{0i_{1}..i_{8}}\partial_{i_{1}}x^{\nu_{1}}..
\partial_{i_{8}}x^{\nu_{8}}\epsilon_{\mu\nu_{1}..\nu_{8}y}
\eeqa
where $\Pi_{\mu}^{(0)}$ is the contribution from $S^{(0)}$
and $\epsilon^{01..8}=1$. So, if the test brane is parallel to the
background brane, a certain orientation lead to the configuration with
1/2 spacetime supersymmetry,
and the other breaks
all the supersymmetry\cite{tsey1}.
The former is reasonable
because there is no force between them.

(f) a test fundamental string in the D-8-brane background

Finally, we discuss a fundamental string in the massive
background. The action is 
\beqa
S=-\int d\xi^{2}\sqrt{-{\rm det}g_{ij}}
+\frac{1}{2}\int d\xi^{2} \varepsilon^{ij}B_{ij}
\label{st-action} 
\eeqa
where the $B_{ij}$ is the worldvolume 2-form induced by
the superspace NS-NS 2-form potential and
$\varepsilon^{01}=1 $.
Since $g_{ij}$ is invariant and $B_{ij}$ is transformed as
$\delta B_{ij}=d(\bar{\varepsilon}^{+}\Delta^{'(1)})$
under (\ref{supertr}), the action is invariant up to total
derivative.
Then, the supercharge is 
$Q_{\alpha}^{+}= Q_{\alpha}^{(0)+}
-i\int_{{\cal M}_{1}}({\cal C}\Delta^{'(1)})_{\alpha}$ where 
$Q_{\alpha}^{(0)+}=\int_{{\cal M}_{1}}d\xi 
[ i\Pi_{\alpha}-\Pi_{\mu}
({\cal C}\Gamma^{\mu}\theta^{+})_{\alpha}]$.
Thus, the superalgebra is obtained as 
\beqa
\{ Q_{\alpha}^{+}, Q_{\beta}^{+}\}&=& 
2\int_{{\cal M}_{1}}d\xi\ \Pi_{\mu}
({\cal C}\Gamma^{\mu})_{\alpha\beta}
-2\int_{{\cal M}_{1}}dy
({\cal C}\Gamma_{y}\Gamma_{11})_{\alpha\beta}
+{\cal O}(\theta^{2}).\label{spalgf1}
\eeqa
The second term means that the orthogonal intersection leads to 
the preservation of 1/4 spacetime supersymmetry, i.e. (0$|$F1,8),
which is deduced from 
(1$|$M2,M9) by a double dimensional reduction of the string
intersection\cite{tow6}.

\section{Discussion}

In summary we have derived the spacetime superalgebras 
from test D-p-branes and a test fundamental string
in a D-8-brane (i.e.{\it a massive IIA})
background, and have shown that the superalgebras
have the same forms as those from the branes not only
in the flat but also in all the other D-p-brane backgrounds
except that half of supercharges are projected out in the
cases of brane backgrounds.
Thus, all the D-brane
solutions including the D-8-brane are on equal footing
from the viewpoint of the superalgebras via
brane probes.
In addition, we have derived from these algebras
all the 1/4-supersymmetric intersections of branes with a D-8-brane
which should be deduced from 
M-brane intersections preserving 1/4
spacetime supersymmetry presented in ref.\cite{tow6}.

As stated in the introduction,
we cannot derive from the superalgebras
bound states which consist of test branes and
the background D-8-brane, 
as opposed to the previoius 
paper\cite{sato1}(in which we have
dealt with the cases of an M-2-brane and an M-5-brane
backgrounds).
This is due to the peculiarity of the harmonic funcion in
the D-8-brane solutions.
The behavior of the harmonic funcions $H_{{\rm p-brane}}$
in p-brane solutions in 10 dimensions
in the limit of vanishing distance is
\beqa
H_{{\rm p-brane}}\to 
\left\{
  \begin{array}{@{\,}ll} 
    \infty & (p=0,1,..,6)\\
    -\infty & (p=7)\\
    0  & (p=8).
  \end{array}
\right.\label{harm}
\eeqa
From this property,
if we take the limit $y\to 0$ in the superalgebras obtained 
in section 3, some inconsistent concequences are drown.
We show this by dealing with the following two cases for example:
a D-2-brane and a D-6-brane 
in the D-8-brane background. 
If the test D-2-brane is parallel to any
2-dimensional subspace of the D-8-brane,
all the supersymmetry is broken.
But the limit $y\to 0$ of the algebra (\ref{spalgd2})
leads to the restoration of 1/2 spacetime supersymmetry
because $\Pi_{\mu}{\cal C}\Gamma^{\mu}\propto H^{1/4}$.
This would be the bound state ``(2$|$2,8)'',
from which we could deduce,
by chain of T-dualities, the bound state (0$|$0,6) preserving 1/2
supersymmetry. This is inconsitent with ref.\cite{pol4}.
In the same way, we could conclude 
from (\ref{spalgd6}) and (\ref{harm})
that (6$|$6,8) preserving 1/2 
supersymmetry would not exist. This configuration, however,
do exist because it can be derived from the bound state
(2$|$M2,M5) preserving 1/2 spacetime
supersymmetry\cite{izq1}\cite{tow5},
by a dimensional
reduction parallel to the M-5-brane but orthogonal to the M-2-brane
and using chain of T-dualities.
Note that we could conclude from the algebras
that the former example (2$|$2,8)
do not exist and that the latter (6$|$6,8) do exist
if the behavior of the harmonic function $H_{{\rm 8-brane}}$
were the same as those in D-p-branes for $p\le 6$.
Thus, we conclude that the inability to deduce the bound states
which consist of branes and a D-8-brane is due to the peculiar
behavior of $H_{{\rm 8-brane}}$ in the limit $y\to 0$.

Related with the above,
it is unclear whether or not the test branes have the boundaries
near the D-8-brane background.
In the cases of M-brane, since proper distances diverge
near the background branes, the ``boundaries'' of the test branes 
are infinitely far away. Thus, they are not considered as their
boundaries and
the test branes are interpreted to have no boundaries, 
as stated in ref.\cite{tow8}.
In this case, however, the proper distance from $y>0$ to
the D-8-brane background is finite,\footnote{
I am thankful to A. Matsuda for the advice on this point.}
which means that the argument presented above 
is not true in this case. 
But since the D-8-brane is a singular surface,
the description of test branes just on it is not thought to be
proper.
So, in this paper we assume that test branes are
within the region of $y>0$ to avoid the singular surface.
We leave this issue for further investigation.

\parbigskipn
{\Large\bf Acknowledgement}
\parbigskipn
I would like to thank Prof. J.Arafune for careful reading of the
manuscript and useful comments. I am grateful to Akira Matsuda for
many helpful discussions and encouragement.
I also would like to thank Taro Tani for useful discussions.

\parbigskipn\parsmallskipn
%
%

\end {document}